\documentstyle[a4,epsfig,11pt,axodraw,amssymb]{article}

\newcommand{\smU}{{\scriptscriptstyle U}}

\newcommand{\sm}[1]{{\scriptscriptstyle #1}}

\begin{document}

\begin{center}
  {\Large\bf Messenger-matter mixing and lepton flavor 
violation} \\
  \medskip
  S.~L.~Dubovsky\footnote{E-mail: \verb|sergd@ms2.inr.ac.ru|},
  D.~S.~Gorbunov\footnote{E-mail: \verb|gorby@ms2.inr.ac.ru|}\\
  \medskip
  {\small
     Physics Department, Moscow State University, 117234
         Moscow, Russia \\
     Institute for Nuclear Research of
         the Russian Academy of Sciences, 117312 Moscow, 
Russia
  }
\end{center}

\begin{abstract}

We consider the Minimal Gauge Mediated Model (MGMM) with
mes\-sen\-ger-matter mixing and find that existing experimental
limits on $\mu \to e\gamma$ decay and $\mu - e$ - conversion
place significant constraints on relevant coupling
constants. On the other hand, the rates of $\tau \to e\gamma$ and
$\tau \to \mu\gamma$ in MGMM are well below the existing
limits. We also point out the possibility of sizeable slepton
oscillations in this model.
\end{abstract}

{\bf1.} Presently, much attention is paid to lepton flavor 
physics in supersymmetric 
theories.
Lepton flavor violation naturally emerges in those SUSY models 
where 
supersymmetry is
broken by supergravity interactions. The corresponding 
soft-breaking terms
are often assumed to be universal at the Planck scale. This
universality breaks down due to the renormalization group 
evolution
below the scale of Grand Unification. In this way one gets
sizeable  mixing in slepton sector at low energies 
\cite{barbieri}. 
This mixing leads to lepton flavor violation for ordinary leptons 
($\mu \to e\gamma$, etc.) at rates close to existing experimental 
limits
\cite{barbieri}, and also to slepton oscillations \cite{krasnikov} 
possibly
observable at the Next Linear Collider.

Another class of SUSY models assumes gauge mediated supersymmetry
breaking \cite{dine}. In these models, the gauge interactions do
not lead to lepton flavor violation because the messenger-matter
interactions are flavor blind.  Nevertheless there is a way to
introduce flavor changing lepton interactions in these
theories. This possibility is based on the observation that some
of the messenger fields have the same quantum numbers as some of
the usual fields. So it becomes natural to introduce direct
mixing between messenger and matter fields \cite{dinem}.  In such
variation of the gauge mediated models, messengers not only
transfer SUSY breaking to usual matter, but also generate lepton
flavor violation.

The purpose of this paper is to show that in a reasonable range
of parameters, messenger-matter mixing in minimal gauge-mediated
models (MGMM) \cite{kolda} gives rise to observable rates of rare lepton
flavor violating processes like $\mu~\to~e\gamma$ and $\mu - e$
conversion.

 We will see that the tree level mixing between leptons is small
due to see-saw type suppression. The tree level mixing between
sleptons is also small in the part of the parameter space that is
natural for MGMM. However, we observe that radiative corrections
involving interactions with ordinary Higgs sector induce much
stronger mixing of sleptons. The point is that messengers obtain
masses not through interactions with ordinary Higgs fields, but
through interactions with hidden sector.  So, the overall matrix
of trilinear couplings is not proportional to the mass matrix
(unlike in the Standard Model). In the basis of eigenstates of
the tree level mass matrix, the largest trilinear terms involve
sleptons as well as messenger and Higgs superfields.  It is these
terms that produce slepton mixing at the one loop level.
 
As a result, at messenger masses of order $10^5$~GeV and Yukawa
couplings of order $10^{-2}$, the rates of $\mu~\to~e\gamma$ and
$\mu - e$ conversion are comparable to their experimental limits.

{\bf2.} The MGMM contains, in addition to MSSM particles, two
messenger multiplets $Q=(q,\psi_q)$ and
$\bar{Q}=(\bar{q},\psi_{\bar{q}})$ which belong to $5$ and
$\bar{5}$ representaions of $SU(5)$.  In what follows we are
interested in lepton sector so we will consider only colorless
components of messenger multiplets. These multiplets couple to a
MSSM singlet $Z$ through the superpotential term
  
\begin{equation}
\label{scalar}
{\cal L}_{ms} = \lambda ZQ \bar{Q}
\end{equation}
$Z$ obtains non-vanishing vacuum expectation values $F$ and $S$
via hidden-sector interactions,
$$Z = S+F\theta \theta $$
Gauginos and scalar particles of MSSM get masses in one and two 
loops 
respectively. 
Their values are \cite{dinem}
\begin{equation}
\label{gaugin}
M_{\lambda_{\sm{i}}} = c_i\frac{\alpha_i}{4\pi}\Lambda f_1(x)
\end{equation}
for gauginos and
\begin{equation}
\label{scalmas}
\tilde{m}^2=2\Lambda ^2\Biggl[ C_3\Big( \frac{\alpha_3}
{4\pi}\Big)^2+
C_2\Big( \frac{\alpha_2}{4\pi}\Big)^2+
\frac{5}{3}\Big( \frac{Y}{2}\Big)^2\Big( \frac{\alpha_1}{4\pi}\Big)
^2\Biggr]f_2
(x)
\end{equation}
for scalars. Here $\alpha_1=\alpha/\cos^2\theta_W$; $c_1=5/3$,
 $c_2=c_3=1$; $C_3=4/3$ for color triplets (zero for singlets),
 $C_2=3/4$ for weak doublets (zero for singlets), $Y$ is the weak
 hypercharge.  The two parameters entering eqs.(\ref{gaugin}) and
 (\ref{scalmas}) are:
$$\Lambda=\frac{F}{S}$$ and 
$$x=\frac{\lambda F}{\lambda^2S^2}$$ 
The dependence of masses on $x$ is very mild, as the functions
$f_1(x)$ and $f_2(x)$ are close to $1$. In the absence of mixing
between messengers and leptons (and/or quarks), the predictions
of this theory at realistic energies are determined predominantly
by the value of $\Lambda$, while the value of $x$ is almost
unimportant.  It has been argued in ref.\cite{borzumati} that
$\Lambda$ must be larger than $70$~TeV, otherwise the theory
would generically be inconsistent with mass limits from LEP.  The
characteristic features of the model are large $\tan{\beta}$ (an
estimate of ref.\cite{borzumati} is $\tan \beta>48$) and large
squark masses \cite{dinem,borzumati}. The next lightest
supersymmetric particle (NLSP) is argued to be a combination of
$\tilde{\tau}_{{\sm R}}$ and $\tilde{\tau}_{{\sm L}}$
\cite{borzumati}. Bino is slightly heavier, but lighter than
$\tilde{\mu}_{{\sm L,\sm R}}$ and $\tilde{e}_{{\sm L,\sm R}}$.

Unlike the masses of MSSM particles, the masses of messenger
fields strongly depend on $x$. Namely, the vacuum expectation
value of $Z$ mixes scalar components of messenger fields and
gives them masses
$$
M_{\pm}^2=\frac{\Lambda^2}{x^2}(1\pm x)
$$ 
It is clear that $x$ must be smaller than $1$.
Masses of fermionic components of messenger superfields
 are all equal to $\frac{\Lambda}{x}$. 

Components of $Q$ have the same quantum numbers as left leptons
(and $\bar{d}$ quarks), so one can introduce messenger-matter
mixing \cite{dinem}
\begin{equation}
\label{y-term}
{\cal L}_{mm} = H_{{\sm D}}L_{\tilde{i}}Y_{\tilde{i}j}E_{j}
\end{equation}
where 
$$H_{\sm D} = (h_{{\sm D}},\chi_{{\sm D}})$$
is one of the Higgs doublet superfields,
\begin{displaymath}
L_{\tilde{i}} = (\tilde{l}_{\tilde{i}},l_{\tilde{i}}) =
 \left\{ \begin{array}{ll}
(\tilde{e}_{{\sm L }\tilde{i}}, e_{{\sm L }\tilde{i}}) & 
\textrm{, $\tilde{i}=1,..,3$}\\
(q, \psi_{q}) & \textrm{, $\tilde{i}=4$}
\end{array} \right.
\end{displaymath}
are left doublet superfields and
$$E_{j} = (\tilde{e}_{{\sm R}j},e_{{\sm R}j})$$
are right lepton singlet superfields.
Hereafter $\tilde{i},\tilde{j}=1,..,4$ label the three left
lepton generations and the messenger field, $i,j=1,..,3$
correspond to the three leptons and $Y_{\tilde{i}j}$ is the
$4\times3$ matrix of Yukawa couplings,
$$
Y_{\tilde{i}j} = 
\left(
\begin{array}{ccc}
Y_{e}& 0 & 0\\
0 & Y_{\mu} & 0\\
0 & 0 & Y_{\tau}\\
Y_{41}& Y_{42}& Y_{43}
\end{array}
\right)
$$
In terms of component fields, the tree level scalar potential 
and
Yukawa terms are
\begin{eqnarray}
\label{scalss}
V & = & \lambda^2S^2 q^*q +
\mu^2h_{{\sm D}}h_{{\sm D}}^* +
 |\lambda S\bar{q} + 
h_{{\sm D}}\tilde{e}_{{\sm R}j}Y_{4j}|^2 + |\mu h_{{\sm U}} + 
Y_{\tilde{i}j}\tilde{l}_{\tilde{i}}\tilde{e}_{{\sm R}j}|^2 
\nonumber \\
& & + |Y_{ij}h_{{\sm D}}\tilde{e}_{{\sm R}j}|^2 + 
|Y_{\tilde{i}j}\tilde{l}_{\tilde{i}}h_{{\sm D}}|^2 +
 \Bigl( \lambda S\psi_{q}\psi_{\bar{q}}
 - \lambda Fq\bar{q} \nonumber \\& & + 
Y_{\tilde{i}j}\bigl( h_{{\sm D}}
l_{\tilde{i}}e_{{\sm R}j} + 
\chi_{{\sm D}}\tilde{l}_{\tilde{i}}
e_{{\sm R}j} + 
\chi_{{\sm D}}l_{\tilde{i}}\tilde{e}_{{\sm R}j}\bigr)
 + \mu\chi_{{\sm D}}\chi_{{\sm U}} + h.c. \Bigr)
\end{eqnarray}
where $\mu$ is the usual parameter of MSSM and
$$H_{\sm U} = (h_{{\sm U}},\chi_{{\sm U}})$$ is the second Higgs
superfield. Besides these terms, there are soft-breaking 
terms coming
from loops involving messenger fields.
In the absence of messenger-matter mixing (\ref{y-term})
 they have the form (at the SUSY breaking scale, which is of
 order of $\Lambda$) 
\begin{equation}
\label{scalsf} 
{\cal L}_{sb} = \tilde{m}_{{\sm L}i}^2\tilde{e}_
{{\sm L}i}\tilde{e}_{{\sm L}i}^* 
+ \tilde{m}_{{\sm R}i}^2 \tilde{e}_{{\sm R}i}\tilde{e}_
{{\sm R}i}^* + B\mu h_{{\sm U}}h_{{\sm D}}
\end{equation}
where $\tilde{m}_{{\sm L}j}^2$ and $\tilde{m}_{{\sm R}j}^2$ are
given by eq.(\ref{scalmas}). Messenger-matter mixing modifies
 eq.(\ref{scalsf}); we will consider this modification later on.
 We will not discuss sneutrinos
in what follows, so we will use the notation 
$e_{{\sm L}j}$ for charged components of lepton doublets.

The main emphasis of this paper is lepton flavor violation
induced by the messenger-matter mixing,
eq.(\ref{y-term}). However, let us note in passing that another
effect of this mixing is the absence of heavy 
stable charged (and
colored) particles (messengers) in the theory \cite{dinem}.

{\bf3.}  To see that lepton mixing is small at the tree level,
let us write the fermion mass matrix (that includes left and
right fermionic messengers) in the following form
\begin{equation}
\label{fermi-d}
{\cal M}_{f} = {\cal U}_{f{\sm L}}{\cal D}_{f}{\cal U}_
{f{\sm R}}
\end{equation}
 where
\begin{equation}
\label{ufl}
{\cal U}_{f{\sm L}} = 
\left(
\begin{array}{cccc}
1& 0& 0& -\frac{y_{e}y_{1}^*}{\lambda^2S^2} \\[2pt]
0& 1& 0& -\frac{y_{\mu}y_{2}^*}{\lambda^2S^2} \\[2pt]
0& 0& 1& -\frac{y_{\tau}y_{3}^*}{\lambda^2S^2} \\[2pt]
\frac{y_{e}y_{1}}{\lambda^2S^2}&  
\frac{y_{\mu}y_{2}}{\lambda^2S^2}&  
\frac{y_{\tau}y_{3}}{\lambda^2S^2}& 1 
\end{array}
\right)
\end{equation}
and
\begin{equation}
\label{ufr}
{\cal U}_{f{\sm R}} = 
\left(
\begin{array}{cccc}
1 - \frac{|y_{1}|^2}{2\lambda^2S^2}& -
\frac{y_{1}^*y_{2}}{2\lambda^2S^2}& 
-\frac{y_{1}^*y_{3}}{2\lambda^2S^2}& -
\frac{y_{1}^*}{\lambda S}\\[2pt]
-\frac{y_{1}y_{2}^*}{2\lambda^2S^2}&
1 - \frac{|y_{2}|^2}{2\lambda^2S^2}&
-\frac{y_{2}^*y_{3}}{2\lambda^2S^2}& -
\frac{y_{2}^*}{\lambda S}\\[2pt]
-\frac{y_{1}y_{3}^*}{2\lambda^2S^2}& 
-\frac{y_{2}y_{3}^*}{2\lambda^2S^2}& 
1 - \frac{|y_{3}|^2}{2\lambda^2S^2}& -
\frac{y_{3}^*}{\lambda S}\\[2pt]
\frac{y_{1}}{\lambda S}& 
\frac{y_{2}}{\lambda S}&
\frac{y_{3}}{\lambda S}& 
1 - \frac{|y_{1}|^2 + |y_{2}|^2 + 
|y_{3}|^2}{2\lambda^2S^2}
\end{array}
\right)
\end{equation}
are mixing matrixes to the leading order in $\frac{y}{\lambda
S}$. Here $v_{\sm U}$ and $v_{\sm D}$ are Higgs expectation
values,
$$y_{i}=Y_{4i} v_{\sm{D}},~~ y_{e,\mu,\tau}=Y_{e,\mu,\tau} 
v_{\sm{D}}$$
and
\begin{equation}
\label{d}
\begin{array}{rcl}
{\cal D}_{f} & = & diag\Bigl(y_{e}\bigl(1 - \frac{y_{e}}
{\lambda S}\bigl| 
\frac{y_{1}}{\lambda S}\bigr|^2\bigr),y_{\mu}\bigl(1 -
 \frac{y_{\mu}}
{\lambda S}\bigl| 
\frac{y_{1}}{\lambda S}\bigr|^2\bigr),
y_{\tau}\bigl(1 - \frac{y_{\tau}}
{\lambda S}\bigl| 
\frac{y_{1}}{\lambda S}\bigr|^2\bigr),\\ & & 
\lambda S\bigl(1 + 
\bigl|\frac{y_{1}y_{e}}{\lambda^2S^2}\bigr|^2 + 
\bigl|\frac{y_{2}y_{\mu}}{\lambda^2S^2}\bigr|^2 +
\bigl|\frac{y_{3}y_{\tau}}{\lambda^2S^2}\bigr|^2\Bigr)
\end{array}
\end{equation}
is the matrix of mass eigenvalues. In principle, the off-diagonal
terms in eqs.(\ref{ufl}) and (\ref{ufr}) may lead to lepton
flavor violation due to one loop diagrams involving sleptons and
gauginos \cite{barbieri}.  However, these mixing terms are
negligible due to see-saw type mechanism: in MGMM one definitly
has $\lambda S>10^4$ GeV, $\tan \beta >1$ and even at $Y_i\sim 1$
the mixing terms are smaller than $10^{-4}$. The corresponding
contributions to lepton flavor violating rates are too small to
be observable.

Mixing in slepton sector is also small at the tree level. 
Indeed, the mass term of scalars
has the following form
$$
{\cal V}_{sc} = s{\cal M}_{sc}s^\dagger
$$
where
\begin{equation}
\label{basis}
s = (\tilde{e}_{{\sm L}},\tilde{\mu}_{{\sm L}},\tilde{\tau}_
{{\sm L}},
\tilde{e}_{{\sm R}}^*,\tilde{\mu}_{{\sm R}}^*,\tilde{\tau}_
{{\sm R}}^*,q,\bar{q}^*)
\end{equation}
are the scalar fields, and 
$$
{\cal M}_{sc}=\left(
\begin{array}{cccccccc}
\tilde{m}_{e_{L}}^2 & 0 & 0 & \mu Y_{e}\upsilon_{\smU}
 & 0 & 0 & y_{e}y_{1}^* & 0\\
0 & \tilde{m}_{\mu_{L}}^2 & 0 & 0 & \mu Y_{\mu}
\upsilon_{\smU} & 0 & 
y_{\tau}y_{2}^* & 0\\
0 & 0 & \tilde{m}_{\tau_{L}}^2 & 0 & 0 &
 \mu Y_{\tau}\upsilon_{\smU} & y_{\tau}y_{3}^* & 0\\
\mu Y_{e}\upsilon_{\smU} & 0 & 0 & \tilde{m}_
{e_{R}}^2 & y_{1}^*y_{2} & y_{1}^*y_{3} & 
\mu Y_{1}^*\upsilon_{\smU} & \lambda Sy_{1}^*\\
0 & \mu Y_{\mu}\upsilon_{\smU} & 0 & y_{1}y_{2}^* &
 \tilde{m}_{\mu_{R}}^2 &
 y_{2}^*y_{3} & 
\mu Y_{2}^*\upsilon_{\smU} & \lambda Sy_{2}^*\\
0 & 0 & \mu Y_{\tau}\upsilon_{\smU} & y_{1}y_{3}^* 
& y_{2}y_{3}^* &
\tilde{m}_{\tau_{R}}^2 & \mu Y_{3}^*\upsilon_{\smU} &
 \lambda Sy_{3}^*\\
y_{e}y_{1} & y_{\mu}y_{2} & y_{\tau}y_{3} & \mu Y_{1}
\upsilon_{\smU} & 
\mu Y_{2}\upsilon_{\smU} & \mu Y_{3}\upsilon_{\smU} & 
\lambda^2S^2 & - \lambda F\\
0 & 0 & 0 & \lambda Sy_{1} & \lambda Sy_{2} & \lambda 
Sy_{3} & 
- \lambda F & \lambda^2S^2
\end{array}
\right)
$$

After the scalar messengers are integrated out at the tree level,
 the lepton flavor violating terms in the mass matrix of right
 sleptons are of order $Y_iY_j(v_{{\sm D}}^2x^2+\frac{\mu ^2
 v_{{\sm U}}^2x^2}{\Lambda ^2})$.  These terms are smaller than
 the one loop contributions (see below) at $\Lambda \gtrsim 10$
 TeV, which is certainly the case in MGMM.  Flavor violating
 mixing between left sleptons is even smaller.  The only
 substantial non-diagonal terms in this matrix are $(- \lambda
 F)$ in the messenger sector and $\tilde{\tau}_{{\sm
 R}}-\tilde{\tau}_{{\sm L}}$ mixing.  Due to the latter, the NLSP
 is likely to be $\tilde{\tau}$ \cite{borzumati}.

{\bf 4.}  The most substantial mixing in slepton sector is due to
one loop diagrams coming from trilinear terms in the
superpotential, that involve $H_{\sm D}$.  The fact that the one
loop mixing terms of sleptons are proportional to the large
parameter $\Lambda^2$ is obvious from eq.(\ref{scalss}): say, one
of the cubic terms in the scalar potential, $[\lambda S
Y_{4j}\bar{q}^*h_{\sm D}\tilde{e}_{{\sm R}j}+$ h.c.], contains
$(\lambda S)=\frac{\Lambda}{x}$ explicitly.

After diagonalizing the messenger mass matrix we obtain the
diagrams contributing to slepton mixing to the order $(\lambda
S)^2$, which are shown in fig.\ref{gopa}.
\begin{center}
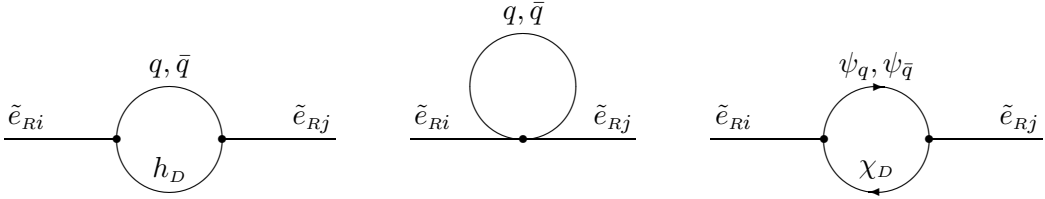
\begin{figure}[htb]
\unitlength=1.00mm
\special{em:linewidth 0.4pt}
\linethickness{0.4pt}
\begin{picture}(148.00,40.00)(12,100)
\put(10.00,120.00){\line(1,0){15.00}}
\put(32.00,120.00){\circle{14.00}}
\put(39.00,120.00){\line(1,0){15.00}}
\put(64.00,120.00){\line(1,0){30.00}}
\put(79.00,127.00){\circle{14.00}}
\put(13.00,122.00){\makebox(0,0)[cb]{$\tilde{e}_{{\sm R}i}$}}
\put(51.00,122.00){\makebox(0,0)[cb]{$\tilde{e}_{{\sm R}j}$}}
\put(67.00,122.00){\makebox(0,0)[cb]{$\tilde{e}_{{\sm R}i}$}}
\put(91.00,122.00){\makebox(0,0)[cb]{$\tilde{e}_{{\sm R}j}$}}
\put(32.00,129.00){\makebox(0,0)[cb]{$q,\bar{q}$}}
\put(32.00,116.00){\makebox(0,0)[cc]{$h_{\sm D}$}}
\put(79.00,136.00){\makebox(0,0)[cb]{$q,\bar{q}$}}
\put(104.00,120.00){\line(1,0){15.00}}
\put(126.00,120.00){\circle{14.00}}
\put(133.00,120.00){\line(1,0){15.00}}
\put(107.00,122.00){\makebox(0,0)[cb]{$\tilde{e}_{{\sm R}i}$}}
\put(145.00,122.00){\makebox(0,0)[cb]{$\tilde{e}_{{\sm R}j}$}}
\put(126.00,129.00){\makebox(0,0)[cb]{$\psi_q,\psi_{\bar{q}}$}}
\put(126.00,116.00){\makebox(0,0)[cc]{$\chi_{\sm D}$}}
\put(126.00,127.00){\vector(1,0){1.00}}
\put(126.00,113.00){\vector(-1,0){1.00}}
\put(79.00,120.00){\circle*{1.00}}
\put(39.00,120.00){\circle*{1.00}}
\put(25.00,120.00){\circle*{1.00}}
\put(119.00,120.00){\circle*{1.00}}
\put(133.00,120.00){\circle*{1.00}}
\end{picture} 
\caption{The diagrams giving main contribution to the scalar
mixing matrix.}
\label{gopa}
\end{figure} 
\end{center}
If supersymmetry were unbroken, their sum would be equal to
zero. In our case of broken supersymmetry the resulting
contribution to the mass matrix of sleptons, to the order
$\Lambda^2$ is
\begin{equation}
\label{delta-mixing}
\delta m^2_{ij}=- \frac{1}{8\pi^2} \frac{\Lambda^2}{x^2} \Big\{ 
-\ln{(1-x^2)}-\frac{x}{2}
 \ln{\Big(\frac{1+x}{1-x}\Big)}\Big\}Y_{4i}^*Y_{4j}
\end{equation}
Since sleptons get negative shifts in $\tilde{m}_{\sm R}^2$, this
equation immediatiatly implies theoretical bounds on Yukawa
couplings $Y_{4i}$ which come from the requirement \cite{dinem}
that none of the slepton masses become negative (see below).

\begin{figure}
{\psfig{file=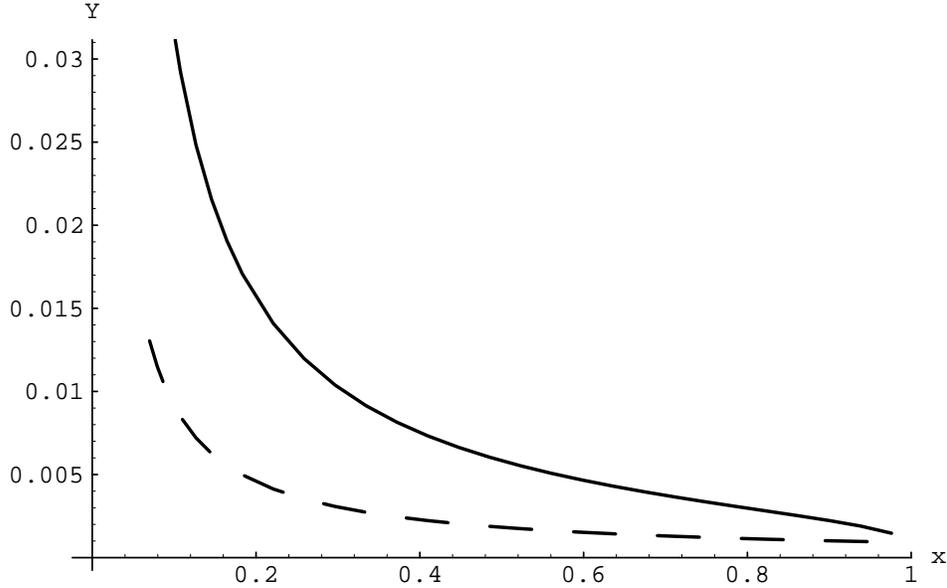,height=8cm,width=12cm}}
\caption
{Upper limit on $Y=|Y_{41}Y_{42}|^{1/2}$ as function of $x$ 
at $\Lambda=100$ TeV (solid line).
Above the dashed line slepton oscillations are unsuppressed at 
maximal mixing}
\label{graphic}
\end{figure}

{\bf 5.}  Let us now consider the effect of slepton mixing on the
usual leptons.  We begin with $\mu \to e\gamma$. The dominant
contributon to the amplitude of this decay comes from diagrams
shown in fig.\ref{popa} where $N_n$ are neutralino mass
eigenstates with masses $M_n$.
\begin{center}
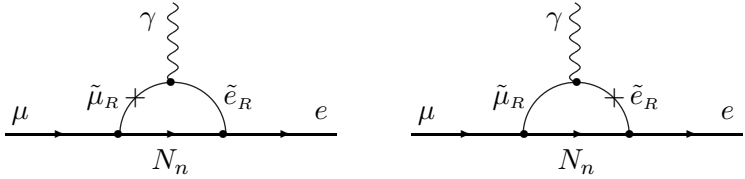
\begin{figure}[htb]
\unitlength=1mm
\linethickness{0.4pt}
\special{em:linewidth 0.4pt}
\begin{picture}(108.00,20.00)(0,120)
\put(10.00,125.00){\line(1,0){44.00}}
\put(17.00,125.00){\vector(1,0){1.00}}
\put(47.00,125.00){\vector(1,0){1.00}}
\CArc(91.50,354.00)(20.0,0,180)
\Photon(91.50,374)(91.50,404){2}{4}
\Photon(243.00,374)(243.00,404){2}{4}
\put(12.00,127.00){\makebox(0,0)[cb]{$\mu$}}
\put(52.00,127.00){\makebox(0,0)[cb]{$e$}}
\put(32.00,123.00){\makebox(0,0)[ct]{$N_n$}}
\put(29.00,140.00){\makebox(0,0)[cc]{$\gamma$}}
\put(23.00,130.00){\makebox(0,0)[cc]{$\tilde{\mu} _{\sm R}$}}
\put(41.00,130.00){\makebox(0,0)[cc]{$\tilde{e}_{\sm R}$}}
\put(64.00,125.00){\line(1,0){44.00}}
\put(71.00,125.00){\vector(1,0){1.00}}
\put(101.00,125.00){\vector(1,0){1.00}}
\CArc(243.50,354.00)(20.0,0,180)
\put(66.00,127.00){\makebox(0,0)[cb]{$\mu$}}
\put(106.00,127.00){\makebox(0,0)[cb]{$e$}}
\put(86.00,123.00){\makebox(0,0)[ct]{$N_n$}}
\put(83.00,140.00){\makebox(0,0)[cc]{$\gamma$}}
\put(77.00,130.00){\makebox(0,0)[cc]{$\tilde{\mu} _{\sm R}$}}
\put(95.00,130.00){\makebox(0,0)[cc]{$\tilde{e}_{\sm R}$}}
\put(32.00,132.00){\circle*{1.00}}
\put(25.00,125.00){\circle*{1.00}}
\put(39.00,125.00){\circle*{1.00}}
\put(79.00,125.00){\circle*{1.00}}
\put(86.00,132.00){\circle*{1.00}}
\put(93.00,125.00){\circle*{1.00}}
\put(27.30,129.80){\makebox(0,0)[cc]{$+$}}
\put(91.00,129.80){\makebox(0,0)[cc]{$+$}}
\put(86.00,125.00){\vector(1,0){1.00}}
\put(32.00,125.00){\vector(1,0){1.00}}
\end{picture}
\caption{Diagrams contributing to  $\mu \to e\gamma$ decay}
\label{popa}
\end{figure}
\end{center}
In general, their conribution gives   
\begin{equation}
\label{me-Gamma}
\Gamma(\mu \to e\gamma)=\frac{\alpha}{4}m_{\mu}^3|F|^2
\end{equation}
\begin{equation}
\label{me-amplitude}
F=\frac{\alpha}{4 \pi cos^2\theta_{{\sm W}}}m_{\mu}
\delta m^2_{12} G(\tilde{m}_{e{\sm R}}^2)
\end{equation}
where 
$$
G(m^2)=\sum_{k=1}^4 \frac{H_{ \tilde{{\sm B}}k}^2}{M_{k}^4} 
g\Big(\frac{m^2}{M_{k}^2}\Big)
$$
$$
g(r)=\frac{1}{6(r-1)^5}\Big(17-9r-9r^2+r^3+6(3r+1)\ln{r}\Big) 
$$
Here $H_{ \tilde{{\sm B}}k}$ are coefficients in the
decomposition of bino in terms of mass eigenstates. To obtain
numerical estimates we point out that neutralino mixing is small,
at least at large mass of bino and large $\mu$.  Neglecting this
mixing, recalling eqs.(2) and (3) and collecting all factors, we
find\footnote{We neglect here the runnung of slepton masses from
the SUSY breaking scale $\Lambda$ down to weak scale. In fact,
this running is small \cite{borzumati}.}
\begin{equation}
\label{rate}
\Gamma(\mu \to e\gamma)=2.6\cdot10^{10}
\frac{m_{\mu}^5}{\Lambda ^4}|Y_{41}Y_{42}|^2\phi (x)
\end{equation}
where
\begin{equation}
\phi(x)=\frac{1}{f_1^8}\left
[g\left(\frac{6f_2}{5f_1^2}\right)\right]^2
\frac{\left[\ln{(1-x^2)}+\frac{x}
{2}\ln\frac{1+x}{1-x}\right]^2}{x^4}
\end{equation} 
and the functions $f_1(x)$ and $f_2(x)$ can be found in
refs. \cite{dimopoulos,martin}.  This rate strongly depends on
$x$, or, in physical terms, on the messenger masses.  In
particular, at small $x$
$$
\phi(x)=\frac{1}{36}\left 
[g\left(\frac{6}{5}\right)\right]^2x^4=
4.2\cdot10^{-5}x^4
$$
To see what eq.(\ref{rate}) means, we plot the limit on the
product of Yukawa couplings as function of $x$ at $\Lambda = 100$
TeV in fig.\ref{graphic}, where we make use of the existing
experimental limit on the rate of $\mu~\to~e\gamma$ decay
\cite{Particle_Data}.
The case of arbitrary $\Lambda$ is straitforward: as follows from
eq.(\ref{rate}), the limit on $|Y_{41}Y_{42}|^{1/2}$ scales like
$\Lambda$ at fixed $x$ and varying $\Lambda$. We conclude that
$\mu~\to~e\gamma$ decay is observable in this model in a
reasonable part of the parameter space.

The slepton mixing in this model gives rise also to $\mu - e$ -
conversion.  The dominant contribution to $\Gamma(\mu \to e\gamma
)$ is given by penguin-type diagrams, while box diagrams are
suppressed by squark masses.  So, there is a simple relation
between $\mu-e$ - conversion and $\mu \to e\gamma$ rates
\cite{barbieri}:
\begin{equation}
\label{meconv}
\Gamma(\mu \to e)=16 \alpha ^4 Z_{eff}^4 
Z|F(q)|^2\Gamma(\mu \to e\gamma)
\end{equation}
For Ti$_{22}^{48}$ with $Z=22$, $Z_{eff}=17.6$, $|F(q)|=0.54$ 
 \cite{titan} one expects
$$
\frac{\Gamma(\mu \to e)}{\Gamma(\mu \to e\gamma)}=
2.8\cdot10^{-2}
$$
while the ratio of experimental limits \cite{collab} is
$$
\frac{\Gamma(\mu \to e)^{lim}_{exp}}
{\Gamma(\mu \to e\gamma)^{lim}_{exp}}=1.1\cdot10^{-1}
$$
Hence, the existing limit on $\mu - e$ - conversion gives weaker
 (by a factor $1.4$) bounds on the product of Yukawa couplings
 $|Y_{41}Y_{42}|^{1/2}$.
\begin{figure}
{\psfig{file=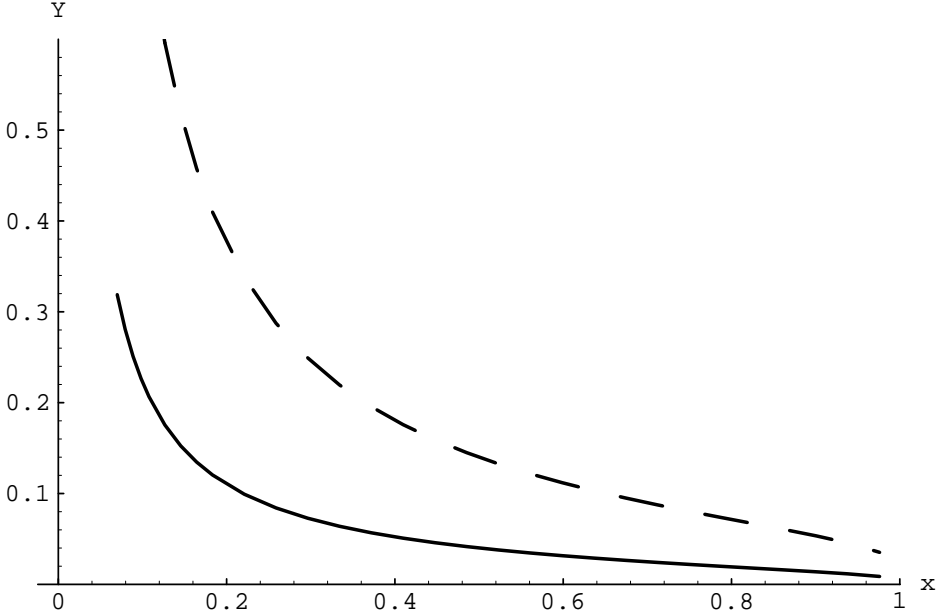,height=8cm,width=12cm}}
\caption
{The upper limits on matrix $Y_{\tilde{i}j}$ from flavor changing
$\tau$ - decays.  The dashed line is the limit on
$|Y_{41}Y_{43}|^{1/2}$ from $\tau \to e\gamma$ decay, the same
line represents also the
limit on $|Y_{42}Y_{43}|^{1/2}$ from $\tau \to \mu\gamma$ decay.
The solid line corresponds to theoretical constraint, 
which is the 
same for $|Y_{41}Y_{43}|^{1/2}$ and $|Y_{42}Y_{43}|^{1/2}$.}
\label{shvah}
\end{figure}

Similar mechanism leads also to flavor changing $\tau$-decays,
$\tau\to e\gamma$ and $\tau\to \mu\gamma$. Crude estimates of the
rates are straightforward to obtain by neglecting
$\tilde{\tau}_{\sm R}- \tilde{\tau}_{\sm L}$ mixing. In this
approximation, $\tau$-decay rates are given by eq.(\ref{rate})
with obvious substitutions $m_{\mu}\to m_{\tau}$, $Y_{42}\to
Y_{43}$ (and $Y_{41}\to Y_{42}$ in the case of $\tau\to
\mu\gamma$). The corresponding upper bounds derived from
experimental limits on $\Gamma (\tau\to \mu\gamma)$ and $\Gamma
(\tau\to e\gamma)$ \cite{Particle_Data} are shown in figure
\ref{shvah}. In fact, these bounds are weaker than theoretical
constraints inherent in this model. Indeed,
 with loop corrections
(\ref{delta-mixing}) to slepton mass matrix included, its
eigenvalues $\tilde{m}_i^2$ are all positive only if
\begin{equation}
\label{ineq}
|Y_{41}|^2+|Y_{42}|^2+|Y_{43}|^2<\frac{5}{3}
\alpha_1^2f_2(x)\frac{x^2}
{\frac{x}{2}\ln\frac{1-x}{1+x}-\ln (1-x^2)}
\end{equation}
Making use of the inequality 
$$|Y_{41}Y_{43}|<\frac{1}{2}(Y_{41}|^2+|Y_{42}|^2+|Y_{43}|^2)$$
one obtains from eq.(\ref{ineq})
theoretical constraint on $|Y_{41}Y_{43}|^\frac{1}{2}$. 
Precisely the same constraint applies to 
$|Y_{42}Y_{43}|^\frac{1}{2}$. The result is shown
in fig.\ref{shvah}.
We see  that self-consistence of the model requires that 
the rates
$\Gamma(\tau \to e\gamma)$ and $\Gamma(\tau \to \mu\gamma)$
 are lower
than the present experimental limits at least by factor 
$10^{-2}$.

{\bf 6.}  Pair production of right sleptons (which decay into
leptons and bino) in $e^+-e^-$ annihilation at the Next Linear
Collider will result in acoplanar $\mu ^+-\mu ^-$ an $e^+-e^-$
events with missing energy.  The NLSP in the model under
discussion is $\tilde{\tau}$, so there will be also four
$\tau$-leptons produced in each event (two $\tau$ will come from
bino decays into $\tau$ and $\tilde{\tau}$ and two more from
subsequent $\tilde{\tau}$ decays). In the presence of slepton
mixing, the slepton oscillations leading to lepton flavor
violating $\mu^{\pm}-e^{\mp}$ events, are possible
\cite{krasnikov}.

Oscillations of sleptons are characterized by the
mixing angle, which in this model can be found
 from eq.(\ref{delta-mixing})
$$
\tan{2\phi}=2\frac{|Y_{41}Y_{42}|}{|Y_{41}|^2-|Y_{42}|^2}
$$
If $Y_{41}\sim Y_{42}$ then mixing is close to maximal.  The
cross section of \linebreak $e^+e^- \to e^{\pm} \mu^{\mp}+4\tau$ may,
however, be suppressed even at large mixing if the lifetime of
$\mu_{{\sm R}}$ and $e_{{\sm R}}$ is small compared to the period
of oscillations. The condition for the absence of such
suppression is \cite{krasnikov}:
\begin{equation}
\label{kras3}
2\Gamma M_{e_{\sm R}} < |M^2_{e_{\sm R}}-M^2_{\mu_{\sm R}}|
\end{equation}
where $M_{e_{\sm R}}$ and $M_{\mu_{\sm R}}$ denote the true
slepton masses.  For slepton decay width $\Gamma$ we have
\begin{equation}
\label{kras4}
\Gamma=\frac{\alpha_1}{2}
\tilde{m}_{_{\sm R}}\bigg(1-\frac{M^2_{bino}}
{\tilde{m}^2_{_{\sm R}}}\bigg)^2
\end{equation}
By making use of eqs.(\ref{gaugin}),(\ref{scalmas}) and 
(\ref{delta-mixing}), 
it is straightforward to 
translate the condition (\ref{kras3}) 
into a condition imposed on Yukawa 
couplings $Y_{41}$ and $Y_{42}$.
For example, at small $x$ one has
\begin{equation}
\label{kras5}
(|Y_{41}|^2+|Y_{42}|^2)x^2 > 1.5\cdot10^{-6}
\end{equation}
In the case of maximal mixing, $Y_{41}=Y_{42}$, the region of
validity of eq.(\ref{kras3}) is shown in fig.\ref{graphic}.
Therefore, there is a fairly wide range of parameters in which
$\mu \to e\gamma$ - decay and slepton oscillations are both
allowed.  Note, that unlike $\mu \to e\gamma$, slepton
oscillation parameters $\sin 2\phi$ and $\frac{2\Gamma M_{e_{{\sm
R}}}}{|M_{e_{{\sm R}}}^2-M_{\mu_{{\sm R}}}|^2}$ are independent
of $\Lambda$.

{\bf 7.}  Messenger-matter mixing is possible also in strongly
interacting sector, as some messengers carry quantum numbers of
right down-quarks. It will lead to flavor changing processes
involving ordinary quarks \cite{dinem}. Yet another possibility
emerges in modifications of the model that are obtained by
changing the messenger sector. For example, one can consider
messengers belonging to $10$ and $\bar{10}$ of $SU(5)$. An
interesting feature of this model is the presence of messengers
with quantum numbers of up-quarks. So one can introduce mixing of
the type $\lambda QLD$ and suggest an interpretation of possible
HERA leptoquark \cite{HERA} as a scalar messenger.  However, to
have its mass of the order of $200$ GeV one has to set $x$ very
close to $1$, which implies fine tuning
$$
\frac{(\lambda S)^2-\lambda F}{(\lambda S)^2}\sim 10^{-6}
$$
We are not aware of any mechanism that would make such fine
tuning natural, but the very possibility of messenger
interpretation of HERA events seems to be interesting.

The authors are indebted V.A.Rubakov for stimulating interest and
numerous helpful discussions. We thank M.V.Libanov and
S.V.Troitsky for valuable criticism. This work is supported in
part by Russian Foundation for Basic Research grant
96-02-17449a.

\end{document}